\documentstyle[mprocl]{article}

\def\Journal#1#2#3#4{{#1} {\bf #2}, #3 (#4)}
 
\def\PRD{{\em Phys. Rev.} D}

\def\be{\begin{equation}}
\def\ee{\end{equation}}
\def\bea{\begin{eqnarray}}
\def\eea{\end{eqnarray}}

\begin{document}
 
\title{RADIATION BACKREACTION IN SPINNING BINARIES}
 
\author{L\'aszl\'o \'A. Gergely, Zolt\'an Perj\'es and M\'aty\'as
Vas\'uth}
 
\address{KFKI Research Institute for Particle
and Nuclear Physics,\\Budapest 114, P.O.Box 49, H-1525
Hungary \\
e-mail: gergely@physx.u-szeged.hu \quad
perjes@rmki.kfki.hu \quad vako@ludens.elte.hu}
 
 
\maketitle\abstracts{ The evolution under radiation backreaction
of a binary system consisting of a black hole and a companion is
studied in the limiting case when the spin of the companion is
negligible compared with the spin $S$ of the black hole. To first
order in the spin, the motion of the reduced-mass particle
excluding radiation effects, is characterized by three constants:
the energy $E$, the magnitude $L$ of the angular momentum and the
projection $L_S$ of the angular momentum along the spin ${\bf S}$.
This motion is quasiperiodic with a period determined by $r_{min}$
and $r_{max}$. We introduce a new parametrization, making the
integration over a period of a generic orbit especially simple.
We give the averaged losses in terms of the 'constants of motion' during
one period for generic orbits, to linear order in spin.
}
 
\section{Introduction}
 
We describe here the gravitational radiation backreaction on two kinds
of binary systems:
a black hole accompanied either by another black hole
with comparable mass (compact binary, {\em CB}) or by
a neutron star viewed as a test particle in the Lense-Thirring ({\em
LT}) picture.
 
 Our recently developed method can be applied
whenever the Newtonian evolution of the binary system is perturbed
in such a way that an uncoupled radial equation exists.
Then for any bounded orbit the turning points $r_{{}_{min}^{max}}$
are at $\dot r=0$. The half period
is defined as the time elapsed between consecutive turning points.
We introduce the {\sl true anomaly parametrization} $r=r(\chi)$
for the evaluation of
integrands of the type: $F/ r^{2+n}$, where $n$ is a positive
integer, defined as:
\begin{eqnarray}
\label{eq:chi}
    \quad  {dr\over d(\cos\chi)} = -(\gamma_0+S\gamma_1)r^2\ ,
    \quad  r(0) = r_{min}\quad
  {\rm and} \quad r(\pi)=r_{max}\
\end{eqnarray}
where $\gamma_0,\gamma_1$ are constants. The integrals over one
period are
conveniently evaluated by computing the residues enclosed in the
circle $\zeta=e^{i\chi}$. This parametrization has the nice
feature that there is only one pole, at $\zeta=0$.
We replace the usual polar angles by new, monotously
changing angle variables.
 
We employ a post-Newtonian $(PN)$ and
additional expansions in both cases:
  In the LT limit an additional expansion over
the small parameter $\eta=m_2/m_1$ is necessary.
For a CB, the spin $S$
is of $PN^{1/2}$ order smaller \cite{ACST} than the orbital angular
momentum $L$.
 
Our main result is that we obtain the leading spin
terms in the {\it averaged} losses of the constants of motion
on {\it generic orbits}.
 
\section{The orbit in the absence of radiation}
 
The equations of motion in the presence of the spin can be derived
from the second order Lagrangian \cite{BOC,TH,KWW}:
\begin{equation}
{\cal L}=\frac{\mu{\bf v}^2}{2}+{g m \mu\over r}+
     {2(1+\eta)g\mu\over c^2r^3}{\bf v}({\bf
r}\times{\bf S})
               +{\eta\mu\over 2c^2m}{\bf v}({\bf a}\times{\bf S})
\end{equation}
where
$r=\vert{\bf r}\vert$ is the relative distance,
${\bf v}$ the relative velocity,
$\mu=m_1 m_2/(m_1+m_2)$ the reduced mass
and $m=m_1+m_2$ the total mass of the
system. The parameter $\eta$ vanishes in the LT case.
 
Up to linear terms in the spin there are three constants of
motion: the energy $E$, the magnitude $L$ and the spin projection
$L_S$ of the orbital angular momentum:
\begin{eqnarray}
\label{eq:E}
E\!\!\!&=\!\!\!&{\mu v^2\over 2}-{g m \mu\over r} + \eta {gL_SS\over
c^2r^3} =
{\frac{\mu}{2}[\dot r^2+r^2(\dot\theta^2+\sin^2\theta\ \dot\varphi^2)]}
-{g m \mu\over r} + \eta {gL_SS\over c^2r^3} \
\label{E}\nonumber\\
\label{eq:L2}
L^2\!\!\!&=\!\!\!&{\mu^2}r^4(\dot\theta^2+\sin^2\theta\ \dot\varphi^2)
-4{g\mu L_SS\over c^2r} + {2\eta\over c^2m}EL_SS \
\label{L2} \\
\label{eq:LS}
L_S\!\!\!&=\!\!\!&{\bf L}\cdot {{\bf S}\over S} = L\cos{\kappa} \ .
\label{LS}\nonumber
\end{eqnarray}
From these a pure radial equation follows:
\begin{equation}
\label{eq:radial}
\dot r^2=2\frac{E}{\mu}+2\frac{gm}{r}-\frac{L^2}{\mu^2r^2}
+2\eta {EL_SS\over c^2m\mu^2r^2}
-2(2+\eta)\frac{gL_SS}{c^2\mu r^3} \ .
\label{radial}
\end{equation}
The true anomaly parametrization, by (\ref{eq:chi}), is:
\begin{eqnarray}
r&=&\frac{L^2}{\mu (gm\mu+A_0\cos\chi)}
-\frac{2\eta L_SS}{c^2m\mu^2A_0}\
\frac{gm\mu^2A_0+(g^2m^2\mu^3+EL^2)\cos\chi}
{(gm\mu+A_0\cos\chi)^2} \\
&+&\frac{2(2+\eta)gL_SS}{c^2L^2A_0}\
\frac{A_0(2g^2m^2\mu^3+EL^2)+gm\mu(2g^2m^2\mu^3+3EL^2)\cos\chi}
{(gm\mu+A_0\cos\chi)^2} \nonumber\
\end{eqnarray}
where $
A_0=(g^2m^2\mu^2+{2EL^2/\mu})^{1/2}$ .
By introducing three Euler angle variables, $\Psi$ (the
argument of the latitude), $\iota_N$ (the inclination of the
orbit) and $\Phi$ (the longitude of the node), the equations of
motion became simpler. From among these angles
only the zeroth order part (in the spin) of the angle $
\Psi=\Psi_0+\chi$
enters the radiative losses.
 
\section{Averaged radiation losses}
 
The instantaneous radiation losses of the constants of motion
are evaluated employing the radiative multipole tensors of
Kidder, Will and Wiseman\cite{KWW}, which originate in the
Blanchet-Damour-Iyer formalism \cite{BDI}. The instantaneous
losses for the energy $E$ and total angular momentum ${\bf J}$
were given by Kidder \cite{Kidder}. From these we
derive both in the LT case and in the CB
case the instantaneous losses of the constants $E,L$ and $L_S$.
 
Now in the averaging process we replace the time integration
by the parameter integration, then we use the residue
theorem to find the averaged radiative losses in the constants of
motion. For instance, the power is
\begin{eqnarray} \label{averloss}
 \Bigl<{dE\over dt}\Bigr>=\!\!\!\!\!&
 -\!\!\!\!\!&{g^2m(-2\mu E)^{3/2}\over15c^5L^7}
 (148E^2L^4+732g^2m^2\mu^3EL^2+425g^4m^4\mu^6)\nonumber\\
\!\!\!\!\!&+\!\!\!\!\!&{S(-2\mu E)^{3/2}g^2\cos\kappa\over10c^7L^{10}}\times\\
\Bigl[520E^3L^6\!\!\!\!\!&+\!\!\!\!\!&10740g^2m^2\mu^3E^2L^4+24990g^4m^4\mu^6EL^2
  +12579g^6m^6\mu^9\nonumber\\
+\eta(256E^3L^6\!\!\!\!\!&+\!\!\!\!\!&6660g^2m^2\mu^3E^2L^4+16660g^4m^4\mu^6EL^2
 +8673g^6m^6\mu^9)\Bigr]\ .\nonumber
\end{eqnarray}
 
The LT limit, $\eta\to0$,
includes the results by Peters and
Mathews \cite{PM,P} for the special case of a
nonspinning black hole; by Shibata \cite{Sh} for an
equatorial orbit about a spinning black hole; and by Ryan
for a circular \cite{Ry1} and generic orbit \cite{Ry2}
about a spinning black hole, respectively.
With the proper definitions of orbital parameters we find perfect
agreement.
 
Complete details of our computations including some subtleties
and the radiative losses in $E$, $L$ and $L_S$
both in terms of the constants of motion and orbital parameters
are described elsewhere \cite{GPV1,GPV2}.
 
\section*{Acknowledgments}
 
  This work has been supported by OTKA no. T17176 and D23744
grants. Financial support for participation at the MG8
Conference for L.\'A.G. from the Unesco and for M.V. from EPS is
acknowledged.

\end{document}